\documentclass[amsmath,amssymb,reqno,tbtags,psamsfonts,10pt,a4paper,twocolumn, prx,superscriptaddress,noeprints]{revtex4-1}

\usepackage{times}

\usepackage[english]{babel}
\usepackage{stmaryrd}
\usepackage{color,graphicx}
\usepackage{upgreek}
\usepackage{dcolumn}% Align table columns on decimal point
\usepackage{bm}% bold math
\usepackage[per-mode=symbol]{siunitx}

\usepackage{array}

\usepackage{cleveref}
\usepackage{csquotes}

\newcommand\eps{\ensuremath{\varepsilon}}

\newcommand{\av}[1]{{\ensuremath{\left\langle #1 \right\rangle}}}

\newcommand{\mig}{\ensuremath_\text{mig}}
\newcommand{\mot}{\ensuremath_\text{mot}}
\newcommand{\DIV}{\ensuremath_\text{div}}

\newcommand{\rss}{\ensuremath{r^\text{ss}_{bf}}}
\newcommand{\VSS}{\ensuremath{v^\text{ss}}}
\newcommand{\Rmax}{\ensuremath{R_\text{max}}}

\newcommand{\dd}{\mathrm{d}}

\newcommand{\cut}{\ensuremath{_\mathrm{cut}}}

\newcommand{\TODO}[1]{}
\begin{document}

\title[]{Control of cell colony growth by contact inhibition}
\bigskip
\author{Simon K. Schnyder}
\email[]{skschnyder@gmail.com}
\affiliation{Fukui Institute for Fundamental Chemistry, Kyoto University, Kyoto, 606-8103, Japan.}

\author{John J. Molina}
\affiliation{Department of Chemical Engineering, Kyoto University, Kyoto 615-8510, Japan.}
\author{Ryoichi Yamamoto}
\affiliation{Department of Chemical Engineering, Kyoto University, Kyoto 615-8510, Japan.}
\affiliation{Institute of Industrial Science, The University of Tokyo, Tokyo 153-8505, Japan.}

\date{\today}% It is always \today, today,

\begin{abstract}
Contact inhibition is a cell property that limits the migration and proliferation of cells in crowded environments. 
However, its role in the emergence of the collective behaviors observed experimentally is not clear.
Here we investigate the growth dynamics of a cell colony composed of migrating and proliferating cells on a substrate using a minimal model that incorporates the mechanisms of contact inhibition of locomotion and proliferation. 
We find two distinct regimes. At early times, when contact inhibition is weak, the colony grows exponentially in time, fully characterized by the proliferation rate. At long times, the colony boundary moves at a constant speed, determined only by the migration speed of a single cell and independent of the proliferation rate. 
Our model illuminates how simple local mechanical interactions give rise to contact inhibition, and from this, how cell colony growth is self-organized and controlled on a local level. 
\end{abstract}

% \pacs{61.43.-j, 64.60.ah, 64.60.Ht, 66.30.H-, 81.05.Rm}
% PACS, the Physics and Astronomy Classification Scheme.

% 61.43.-j	Disordered solids (see also 81.05.Gc Amorphous semiconductors, 81.05.Kf Glasses, and 81.05.Rm Porous materials; granular materials in materials science; for photoluminescence of disordered solids, see 78.55.Mb and 78.55.Qr)
% 64.60.ah	Percolation
% 64.60.Ht	Dynamic critical phenomena (for quantum critical phenomena in superconductivity, see 74.40.Kb)
% 66.30.-h	Diffusion in solids (for surface and interface diffusion, see 68.35.Fx)
% 66.30.H-	Self-diffusion and ionic conduction in nonmetals
% 81.05.Rm	Porous materials; granular materials (for granular superconductors, see 74.81.Bd)
% 82.70.Dd	Colloids

\maketitle

\section{Introduction}

Cells move collectively
 and proliferate~\cite{Friedl2009,Rorth2009,Angelini2011,Mehes2014,Mayor2016,Hakim2017} 
 as the embryo develops during morphogenesis~\cite{Lecuit2007}, 
as cancer spreads or as wounds close~\cite{Poujade2007,Trepat2009,Petitjean2010}. The way in which migration and proliferation interact with each other is complex~\cite{Gauquelin2019}. %"Collective migration of an epithelial monolayer in response to a model wound" and "Physical forces during collective cell migration" 
Essential for the regulation of these processes is contact inhibition of locomotion (CIL), 
which describes the tendency of cells to stop migration or change direction when coming into contact with other cells~\cite{Abercrombie1953,Abercrombie1979,Carmona-Fontaine2008,Scarpa2015,Stramer2016}. 
CIL has been shown to enable cells to collectively migrate \cite{Carmona-Fontaine2008}, follow chemical gradients~\cite{Theveneau2010,Carmona-Fontaine2011}, and disperse. It is now believed that CIL helps control of collective tissue migration~\cite{Stramer2016,Theveneau2010,Coburn2013,Smeets2016}, tissue growth~\cite{Puliafito2012,Zimmermann2016}, morphogenesis \cite{Carmona-Fontaine2008}, wound healing and tumour development~\cite{Li2014}.

Potentially distinct from CIL is contact inhibition of proliferation (CIP) which refers to the suppression of cell divisions in dense regions of tissues \cite{Fisher1967,Fujito2005,Takai2008}, which in turn regulates their growth. There is evidence that CIP does not require direct cell contact \cite{Stoker1973,Stoker1974,Dunn1984,MartzE1972} and as a consequence the effect is also called “density dependent inhibition of cell growth”~\cite{Stoker1967}.

Modelling approaches for cell migration are manifold~\cite{Camley2017}, 
ranging from strongly idealised models for single cells crawling on substrates \cite{Tarama2018,Goychuk2018, Molina2018} to cells in confluent tissues \cite{Bi2015} to continuum theories \cite{Blanch-Mercader2014,Yabunaka2017,Yabunaka2017b}).
The collective behavior of cells is under intense study and many questions about contact inhibition are still open \cite{Takai2008,Stramer2016}. 

In order to reduce the complexity of the systems under study, it is valuable to investigate well controlled model systems. One such model system deals with the crawling and proliferation of a monolayer of cells seeded onto a substrate. How a few cells develop into an extended colony has been investigated in a recent experiment performed by \citet{Puliafito2012}, while numerical approaches were for instance given by \citet{Drasdo1995,Radszuweit2009,Basan2011,Basan2013,Farrell2013,Farrell2014,Zimmermann2016}. 
\citet{Puliafito2012} observed two regimes in their investigation of a colony of Madin-Darby canine kidney epithelial (MDCK) cells~\cite{Gaush1966}. In the beginning the colony's area grows exponentially with time, followed by subexponential growth. In the former regime, cells are highly mobile and divide frequently, while in the latter, both the motion and proliferation of the cells becomes suppressed, linking the transition to contact inhibition.

Previously, we developed a minimal, mechanical model for crawling cells in which contact inhibition of locomotion arises from the internal mechanics of the cells~\cite{Schnyder2017}. Our model focuses purely on the cell mechanics, since mechanical forces inside of and between cells are now understood to be of crucial importance for the understanding of cell dynamics~\cite{Xi2018,Roca-Cusachs2017,Yang2017,Shraiman2005,SerraPicamal2012,Trepat2009,Tambe2011,Puliafito2012}.
 We found that these model cells naturally exhibit a range of realistic behaviors, including the emergence of collective migration. Our model is thus a natural candidate to investigate colony growth.

In this paper, we extend our model to include cell proliferation in such a way that the motility and division dynamics are entirely governed by the internal dynamics of the cells. The cells cycle between a motile phase and a division state. In the division state, the cells attempt to proliferate by making space for two cells; otherwise, the cell division is aborted. This naturally gives rise to a form of contact inhibition of proliferation.

We found that our model reproduces the typical colony dynamics; with exponential growth at short times turning into subexponential growth with a constant boundary speed at long times. Coinciding with this transition, the average cell speed decreases strongly, because of CIL occurring in the inside of the colony. As a result of contact inhibition, cells close to the boundary have higher speeds and proliferation rates. We identify simple scaling relations for both regimes and the crossover between them.
We had previously found that cell shape has a strong effect on cell collisions and that cells with large front disks align and coherently migrate~\cite{Schnyder2017}. In this work, we now see that cell with large fronts orient themselves away from the colony, which enhances the speed of colony expansion.

\section{Model}

\begin{figure}
  \centering
    \includegraphics[width=\columnwidth]{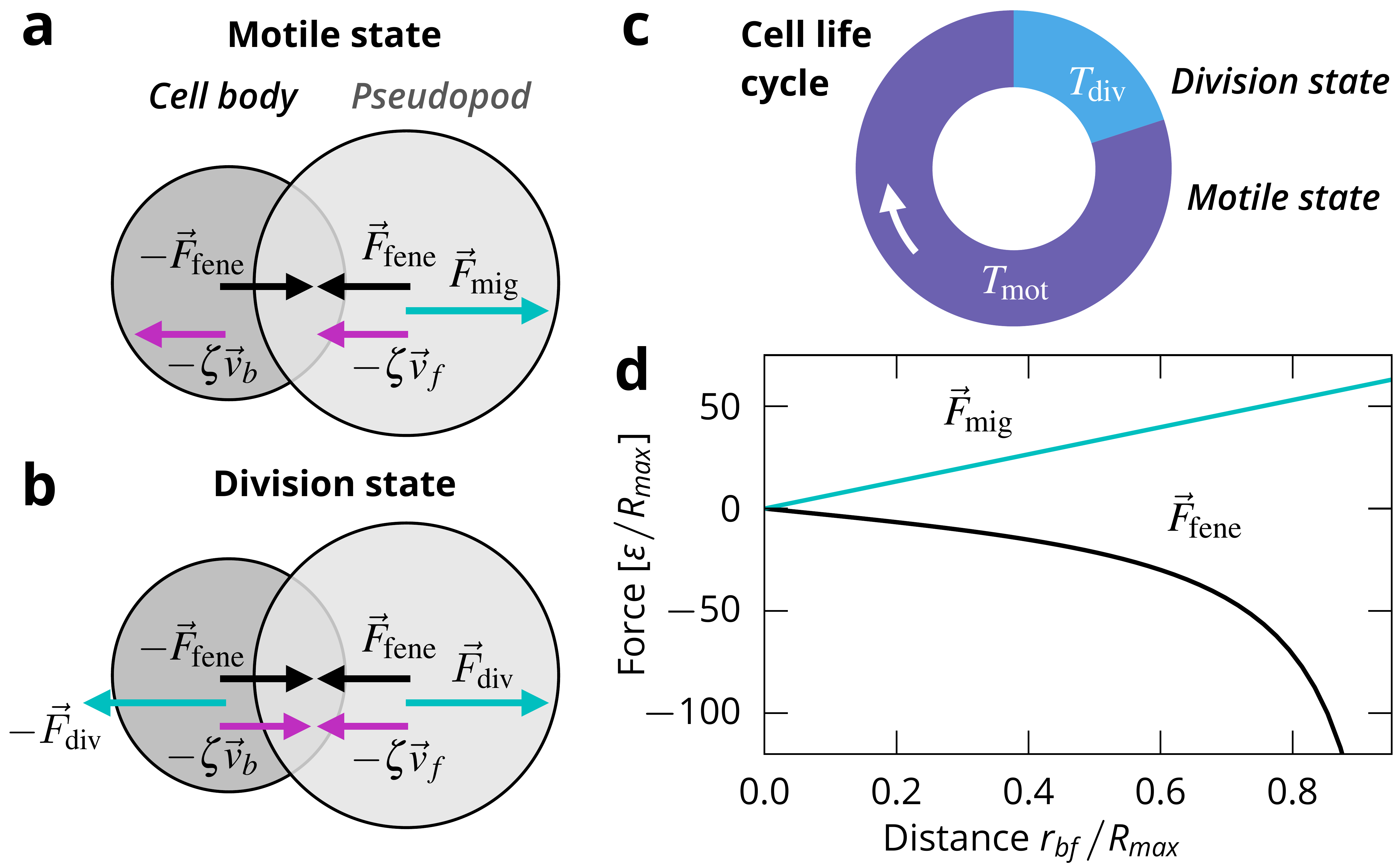}
  \caption{\textbf{Model overview. }(a) Schematic of the cell model in the motile state. 
  (b) Schematic of the cell model in the division state 
  (c) Illustration of the cell life cycle.
  (d) Forces acting on the disks being apart a distance $r_{bf} = |\vec r_{bf}|$.
	}
  \label{fig:CellMigrationCycle}
  \label{fig:forces}
\end{figure}

We build on a model for crawling cells~\cite{Schnyder2017}, and include cell division. Each cell consists of two disks with distinct roles, see \cref{fig:CellMigrationCycle}. One models a pseudopod and is at the front of the cell (index $f$), the other disk represents the cell body and is at the back of the cell (index $b$). The positions of the disks $r_f$ and $r_b$ define the distance between the elements and the orientation of the cell $\vec r_{bf} = \vec r_f - \vec r_b$.

The dynamics of the cells are coarse-grained over the typical idealized crawling cycle \cite{Ananthakrishnan2007}. The substrate exerts a drag force $-\zeta_i \vec v_i$ on the disks, with $\vec v_i$ being the velocity of the disks ($i \in {f,b}$) and $\zeta_i$ being the respective drag coefficients. For simplicity, we set $\zeta_1$ = $\zeta_2$ = $\zeta$. We neglect intracellular friction and friction between different cells, assuming that the friction with the substrate is the dominant contribution to friction in the system.

The cell shape is determined by the configuration of the two disks with diameters $\sigma_f$ and $\sigma_b$, and can be understood as representing a statistical average over the real cell shape. It is a coarse graining of the highly variable shape of real cells. This is a promising approach for cells whose shape does not deviate not too much from the average, e.g. for epithelial cells.

In our previous work, interactions between cells were purely repulsive. Here, we introduce new, adhesive interactions between cells. The adhesiveness of the potential is characterised by its well depth $\eps_\text{well}$ in respect to the potential height $\eps_\text{core}$. 
The force acting on a disk $\alpha$ by other cells is denoted $F_{cc, \alpha}$ ($\alpha \in [b, f]$). For details, see Methods. 

\subsection{Cell life cycle}

Each cell switches independently between two states, a motile state and a division state. The duration of the motile state is determined when switching to it from the division state, by drawing a random number from a normal distribution with mean $T\mot$ and standard deviation $T\mot/2$. The duration of the division state is held constant at $T_\text{div}$. On average, the duration of the whole cell cycle is thus $T = T_\text{div} + T\mot$.

In the motile state, the cell behaves as in Ref.\ \citenum{Schnyder2017}.
The front disk exerts a migration force $F\mig$ in the direction of the orientation of the particle, while the back disk is passive. The migration force is given by
	$\vec F_\text{mig} (\vec r_{bf}) = m \vec r_{bf}$
with motility strength $m$, see \cref{fig:forces} (a,d).
The connection between the disks is modelled as a finitely extensible nonlinear elastic (FENE) spring \cite{Jin2007}, which gives rise to a contracting force between the disks
	$\vec F_\text{fene}(\vec r_{bf}) = - \kappa \vec r_{bf}/[1-(r_{bf}/R_\text{max})^2]$
with coupling parameter $\kappa$ determining the strength of the contraction, and $R_\text{max}$ the maximum distance between the two disks.

In the division state, cells attempt to make space for two daughter cells. Cells only divide if at the end of the division state the cell extension reaches a division threshold $R_\text{div}$. Coupling cell proliferation to cell area is an idealization of the observation that larger cells divide far more frequently in experiments \cite{Puliafito2012} and models contact inhibition of proliferation.

Cells elongate by having both disks enact the same migration force $F_\text{div}$ with opposite sign. For convenience we set $F_\text{div} = F_\text{mig}$. The contracting force between the disks remains unchanged.
Cells do not migrate in the division state. To make space for the new cells, we increase the size of the smaller of the two disks linearly until it matches the size of the larger disk at the end of the state.
After a successful division we construct two cells in place of the original cell's disks. We randomly displace the new cell's disks to randomize the orientation of the new cells. 
If $r_{bf} < R_\text{div}$ at the end of the division state, the cell division is aborted, the cell contracts again, and the migration state is entered.

\subsection{Equations of motion}

For each of the cells, we now have two coupled non-linear equations of motion, assuming overdamped dynamics
\begin{gather}
	\begin{aligned}
		\label{eq:equationsOfMotion}
		\frac{\dd}{\dd t} \vec r_b &= \frac{1}{\zeta} \left(- \vec F_\text{fene} (\vec r_{bf}) - \chi \vec F_\text{mig} (\vec r_{bf}) + \sum_\text{neigh.} \vec F_\text{cc,b} \right),\\
		\frac{\dd}{\dd t} \vec r_f &= \frac{1}{\zeta} \left(\vec F_\text{fene} (\vec r_{bf}) +  \vec F_\text{mig} (\vec r_{bf}) + \sum_\text{neigh.} \vec F_\text{cc,f} \right),
	\end{aligned}
\end{gather}
with $\chi = 1$ in the division state and $\chi=0$ in the motile state.

Apart from the randomized positions of daughter cells' disks, our model does not include random forces. This is a reasonable assumption when collisions (and cell division) dominate the dynamics \cite{Drescher2011,Wensink2012}. 
In the migration state, the cell is only motile when its disks have some separation, $r_{bf}>0$, and thus when its shape deviates from a circle. This kind of coupling of motility and deformation is typical in migratory cells \cite{Nelson2009}.

If the cell is in the motile state for long enough, it then enters a steady state with constant extension $\rss$  and constant speed $\VSS$ in which the forces acting on the cell balance \cite{Schnyder2017}.

\section{Results}

\subsection{Colony growth}

\begin{figure}
  \centering    \includegraphics[width=\columnwidth]{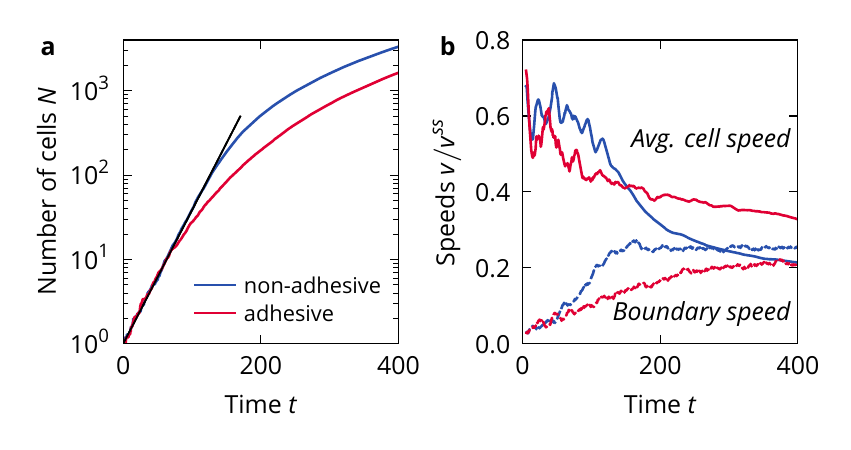}
  \caption{\textbf{Colony growth for non-adhesive and adhesive cells.} (a) Size of the colony in number of particles $N$ against time $t$ for $T\mot = 16$ and $T\DIV = 3$. The black line marks exponential growth, assuming that all cell division attempts are successful, see \cref{eq:exponentialAsymptote}. (b) Speed of the colony boundary and average cell speed of the same simulations. }
  \label{fig:colonyGrowth_velocity_vs_time}
  \label{fig:velocity_vs_time}
\end{figure}

At first, we simulated cell colonies of non-adhesive cells.
At early times, the colony grows exponentially, but eventually crosses over into subexponential growth, see \cref{fig:colonyGrowth_velocity_vs_time}a). 
In the exponential regime, all cell division attempts are successful. Since cells attempt to double with a rate of $T = T_\text{mig} + T_\text{div}$ and we always start with one cell at $t = 0$, the number of cells grows as 
\begin{align}
  N(t) = 2^{t/T} = \exp\left(\frac{\ln 2}{T} t \right).
  \label{eq:exponentialAsymptote}
\end{align}
In the experiment by \citet{Puliafito2012}, the subexponential growth is characterized by the boundary of the colony moving outwards at a constant speed.
If $R(t)$ is the radius of the approximately circular colony, then the outwards speed of the boundary can be extracted from the area $A(t) = \pi R(t)^2$ of the colony as
\begin{align}
	v_B  = \frac{\dd R(t)}{\dd t} = \frac{\dd}{\dd t} \sqrt{A(t)/\pi}.
  \label{eq:boundarySpeed}
\end{align}
The area of the colony can be calculated from the areas of all the individual cells, for which we use \cref{eq:cellArea}.
At long times, the speed of the boundary saturates in our simulation to a constant speed as well, see \cref{fig:colonyGrowth_velocity_vs_time}b). We find the speed to be $v_B\approx 0.2\VSS$. 
The speed of cells allows quantifying the activity of the colony over time. In the exponential regime, the average cell speed is $\av {v} \approx 0.6\VSS$ and then decreases over time, eventually dropping below the boundary speed, see \cref{fig:colonyGrowth_velocity_vs_time}b). The transition in the average cell speed occurs at the same time as the transition of the boundary speed. All of this is qualitatively similar to what is observed in the experiment \cite{Puliafito2012}.

Adhesive cells exhibit similar growth dynamics, see \cref{fig:colonyGrowth_velocity_vs_time}. The slope of the exponential regime is the same, with all divisions being successful at early times. The average cell speed in the exponential regime remains unchanged as well.
However, the exponential regime only extends until the colony consists of tens of cells. The transition to a constant boundary speed takes much longer, as does the slowing of the average cell speed. On average, we find cells to be faster, but the colony to expand slower. 

\subsection{Radial analysis}

\begin{figure}
  \centering   
  \includegraphics[width=\columnwidth]{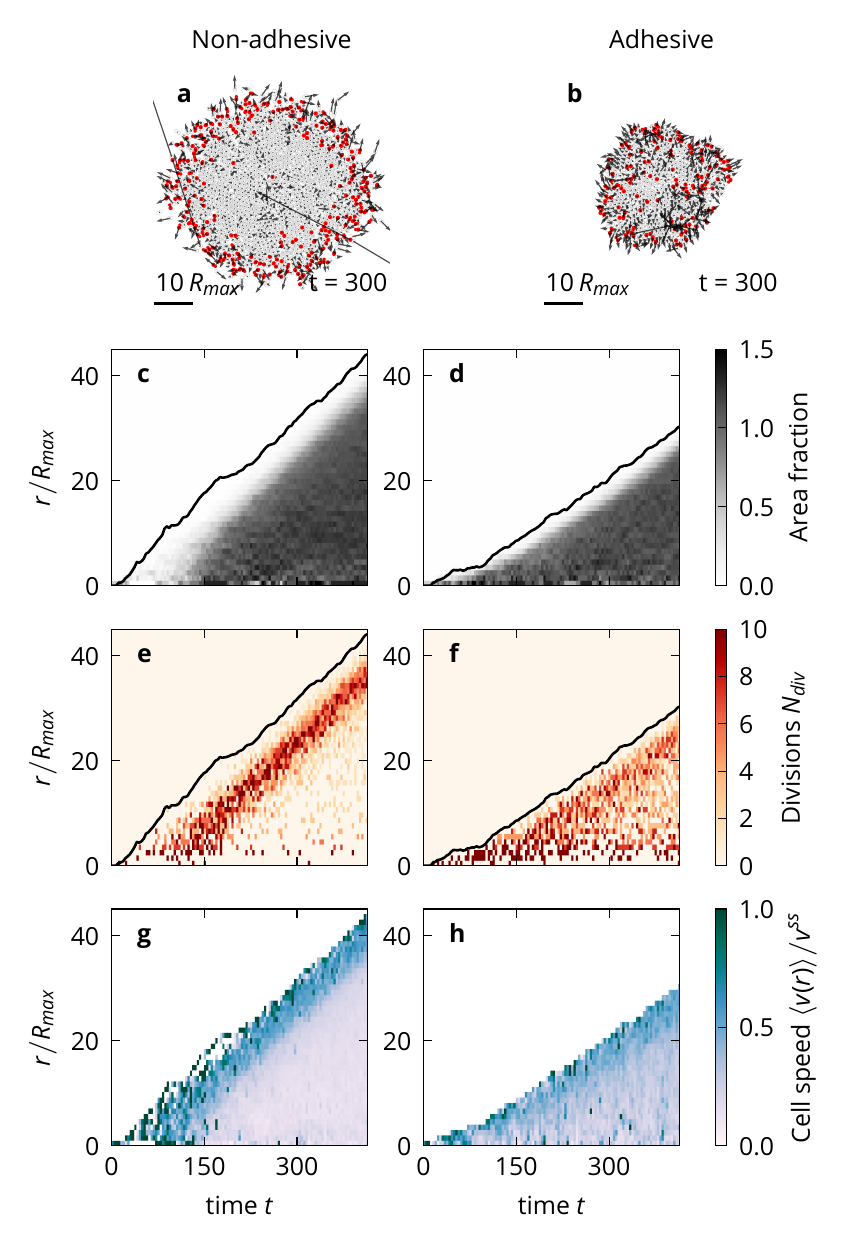}
  \caption{\textbf{Radial analysis.} (a) Simulation snapshot of a colony of non-adhesive cells. Cells marked in red were created via cell proliferation in the preceding time span $\Delta t = 20$. Cell speeds are given as arrows (b) The same as (a) but for adhesive cells. 
  (c) Cell density for the non-adhesive cell simulation as a function of distance to the colony center and time. The distance of the outermost cell for each point in time is indicated by the black line. (d) The same as (c) but for adhesive cells. 
  (e) Number of cell divisions per unit time for the non-adhesive cell simulation as a function of distance to the colony center and time. (f) The same as (e) but for adhesive cells
  (g) Average cell speed for the non-adhesive cell simulation as a function of the distance from the colony center and time. (h) The same as (g) but for adhesive cells.}
  \label{fig:colonyGrowth_kymos}
\end{figure}

To understand the growth of the colonies in more detail, we look at the spatial distribution of the following key quantities: cell density, cell divisions, and cell speed. For this we analyse one exemplary simulation with non-adhesive and adhesive cells.
Non-adhesive cells form colonies with a diffuse boundary, with some cells even escaping, \cref{fig:colonyGrowth_kymos}a), whereas adhesive cells form a denser colony, with no cells escaping the bulk, see \cref{fig:colonyGrowth_kymos}b).

For the non-adhesive cells, we find that in the exponential regime ($t<150$) the density is quite low but later quickly reaches a high value in the bulk of the colony, \cref{fig:colonyGrowth_kymos}c). The boundary of the colony remains diffuse with a low density and constant width over the whole simulation. The colony of adhesive cells is different at early times, see \cref{fig:colonyGrowth_kymos}d), because it is dense and cohesive from the beginning. At long times, the adhesive colony appears qualitatively quite similar to the non-adhesive colony, with the same bulk density, albeit with a sharper boundary.

We then calculated the radial distribution of successful cell divisions $N_\text{div}(r, t)$, see Methods. In the exponential regime, all attempted divisions are successful, and thus are distributed homogeneously over the colony, \cref{fig:colonyGrowth_kymos}e,f). For non-adhesive cells, divisions at later times mostly occur in a ring of constant thickness at the boundary of the colony, where there is more space available to the cells, see \cref{fig:colonyGrowth_kymos}a,e). 
This is a result of our cell division mechanism which naturally gives rise to contact inhibition of proliferation. 
Cells in the bulk cannot make the necessary space for the cell division to occur, except for the colony center, where space becomes available as cells migrate outwards.
For adhesive cells we find a similar pattern, see \cref{fig:colonyGrowth_kymos}f), where a ring of increased proliferation probability can still be discerned.
However, the probability of a division occurring at the boundary is considerably reduced, because the local density tends to be higher as compared to the non-adhesive colony. In addition, cell divisions are more frequent in the bulk of the colony. The reason for this becomes clearer with an analysis of the cell speeds.

Non-adhesive cells are mostly mobile at the boundary of the colony, with motion being strongly suppressed by CIL in the colony bulk, see \cref{fig:colonyGrowth_kymos}a). 
The cells on the boundary are on average pointing away from the colony center, but there is considerable local variation due to contact inhibition of locomotion after collisions and noise introduced by cell divisions. Cells can momentarily obtain high speeds after a successful cell division when they move away from the other daughter cell.

In the exponential regime, all cells are mobile, but in the subexponential regime, only cells at the boundary exhibit speeds close to the single-cell steady state speed $\VSS$, \cref{fig:colonyGrowth_kymos}g). Motion in the bulk is suppressed strongly by contact inhibition of locomotion. 
In comparison, adhesive cells are slower on the border but more mobile in the bulk and more aligned with each other, see \cref{fig:colonyGrowth_kymos}b, h). 
We attribute this to the cells at the boundary being held back by the cells at their back and in turn the cells of the bulk being pulled outwards by the boundary cells. This is commonly called the "tug of war" between cells \cite{Trepat2009,SerraPicamal2012,Tanimoto2012}.
The tug of war leads to a different bulk structure between the colonies, even though the densities are similar, \cref{fig:colonyGrowth_kymos}a,b). There is more free space in the non-adhesive colony and cells are more compressed due to contact inhibition of locomotion. The adhesive colony, on the other hand, is fully cohesive, with cells always being at contact and pulling on each other and therefore, on average, more extended. This is reflected in the higher cell speeds in the bulk of the adhesive colony. As a result, it becomes easier for the adhesive cells in the bulk to reach the division threshold. In conclusion, we find that cell-cell adhesion considerably alters the colony structure on a local level, while leaving the qualitative colony dynamics with the two growth regimes unchanged. 

\subsection{Variation in the cell cycle}

\begin{figure}
  \centering
    \includegraphics[width=\columnwidth]{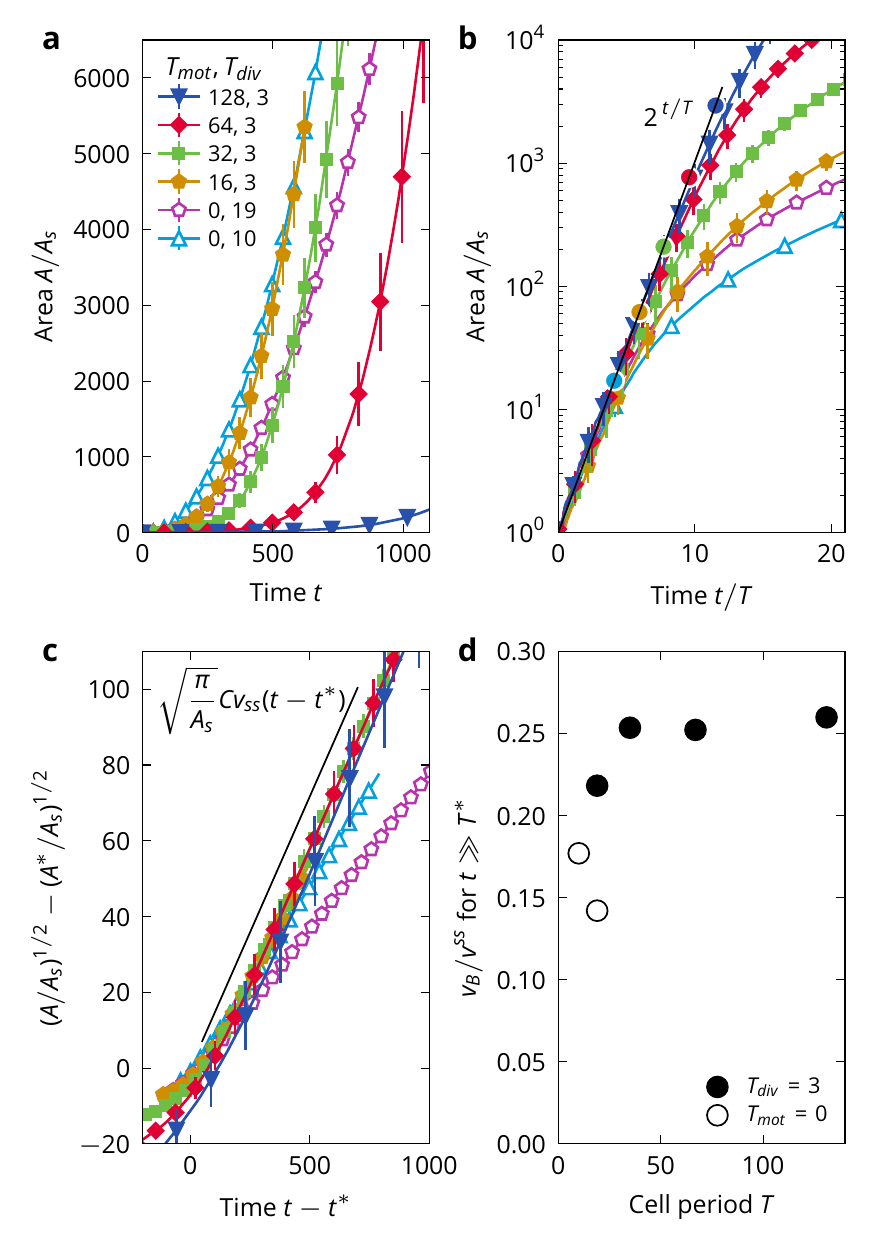}
  \caption{\textbf{Variation in the cell cycle.}
  a) Area of the colony for a range of $T\mot$ against time. Error bars calculated from the standard deviation of colony sizes for the independent simulation runs are marked as vertical lines at regular intervals.
  b) Area of the colony for a range of $T\mot$ against rescaled time. The black line marks the exponential growth expected for when all cell division attempts are successful, \cref{eq:shortTimeScaling}. The colored circles mark the estimates for the crossover time $t^*$ and crossover colony size $A^*$, as determined from \cref{eq:Tstar,eq:Astar}. The simulations with $T\mot, T_\text{div} = 16,3$ and $0,19$ share the same crossover.
  c) Scaling plot to expose linear long-time growth. The black line gives the asymptote of \cref{eq:longTimeScaling}.
  d) Boundary speed against the cell cycle period $T$.
	}
  \label{fig:cg_Tmot_scaling}
  \label{fig:cg_Tmot_subexponential}
\end{figure}

We illustrate in the following that at short times, the colony dynamics is entirely determined by the proliferation rate, whereas at long times, the dynamics are determined by the migration properties of the cells. For this purpose we vary the durations of the migration and division states of the cells, \cref{fig:cg_Tmot_scaling}a). The shortest tested cell cycle is with $T\mot = 16$. Compared to the migration timescale of $\tau_\text{mig} = 5.33$ which a solitary cell would need to migrate roughly it's own length, this is quite short and marks the lower end of applicability for our model. 
The longest cell cycle is $T\mot = 128$, which is over an order of magnitude larger than the migration time scale and thus comparable to MDCK cells. 
Finally, for comparison, we also simulated the situation where the cells are not migrating at all, but divide regularly.

Assuming that the colony area grows as 
\begin{align}
  A(t) \approx A_s N(t) = A_s 2^{t/T}
  \label{eq:shortTimeScaling}
\end{align}
at early times, we find that all simulations collapse onto one asymptote when time is rescaled by $T = T\mot + T_\text{div}$, see \cref{fig:cg_Tmot_scaling}b). Therefore, while contact inhibition of proliferation is weak, the dynamics are purely determined by the rate of cell divisions $1/T$. As a consequence, it also does not play a role  whether the cells are migratory or not.

However, this simple rescaling cannot collapse the data at long times, as the time at which growth becomes subexponential varies between the simulations. To calculate those crossover times, we make use of the observation that the boundary speed of the colonies is limited. In the exponential regime, the boundary speed increases as, see \cref{eq:boundarySpeed,eq:shortTimeScaling},
\begin{align}
	v_B(t) = \frac{\dd }{\dd t} \sqrt{ \frac{A(t)}{\pi}} \approx \frac{\ln 2}{2T}\sqrt{\frac{A_s}{\pi}} 2^{t/2T}.
\end{align}
The crossover time $t^*$ is reached when the colony reaches the terminal boundary velocity which we assume to be proportional to the steady state speed of a single cells, $v_B^* = C \VSS$ with some factor $C$. Then we obtain
\begin{align}
  t^* = T \log_2\left[\frac{4\pi}{(\ln 2)^2}\frac{(C\VSS T)^2}{A_s}\right]
  \label{eq:Tstar}
\end{align}
From the data, we find that $C = 0.25$ holds generally for migrating cells, see \cref{fig:cg_Tmot_scaling}d). 
 With this value for $C$, the crossover times correctly mark the transitions to sub-exponential growth for all simulations, see  \cref{fig:cg_Tmot_scaling}b). From \cref{eq:Tstar}, we see that the crossover is most strongly influenced by the duration of the cell cycle, followed by the crawling speed of the cells.
At the crossover, the colony is of size $A^* = A_s 2^{t^*/T}$. This is equivalent to 
\begin{align}
  t^* = T \log_2 (A^*/A_s).
\end{align}
and thus
\begin{align}
  \label{eq:Astar}
  A^* = \frac{4\pi}{(\ln 2)^2}{(C\VSS T)^2}\text{ and }
  R^* = \frac{2}{\ln 2} C \VSS T
\end{align}
At long times, $t \gg t^*$, the radius of the colony grows with constant speed $C \VSS$, so we find that approximately it must hold that
\begin{align}
  R(t) \approx R^* + C \VSS (t-t^*).
\end{align}
Rewritten for the colony area, we have
\begin{align}
  \sqrt{A(t)} - \sqrt{A^*} = \sqrt{\pi} C \VSS (t-t^*)
  \label{eq:longTimeScaling}
\end{align}
This long-time scaling is exposed in \cref{fig:cg_Tmot_scaling}c). Most of the data collapse onto a master function that is close to \cref{eq:longTimeScaling}. The two simulations that deviate from this are the two cases in which the cells do not actively migrate and thus necessarily violate the scaling. 

We therefore find that if the cells are actively migrating, the colony boundary moves at a constant speed determined by cell motility, and that if the cells are not migrating, long time growth becomes severely suppressed.
Or stated differently, regardless of the proliferation rate, at long times the colony expands as fast as the cells on the boundary are able to migrate away from the colony center.

Notably, the boundary speed is smaller than the steady state speed of a solitary cell $\VSS$ by the factor C. This has two reasons: (1) the cells on the boundary adhere to the rest of the colony and are pulled back by them 
(2) The cells are not perfect at orienting themselves away from the colony.

\begin{figure}
  \centering
    \includegraphics{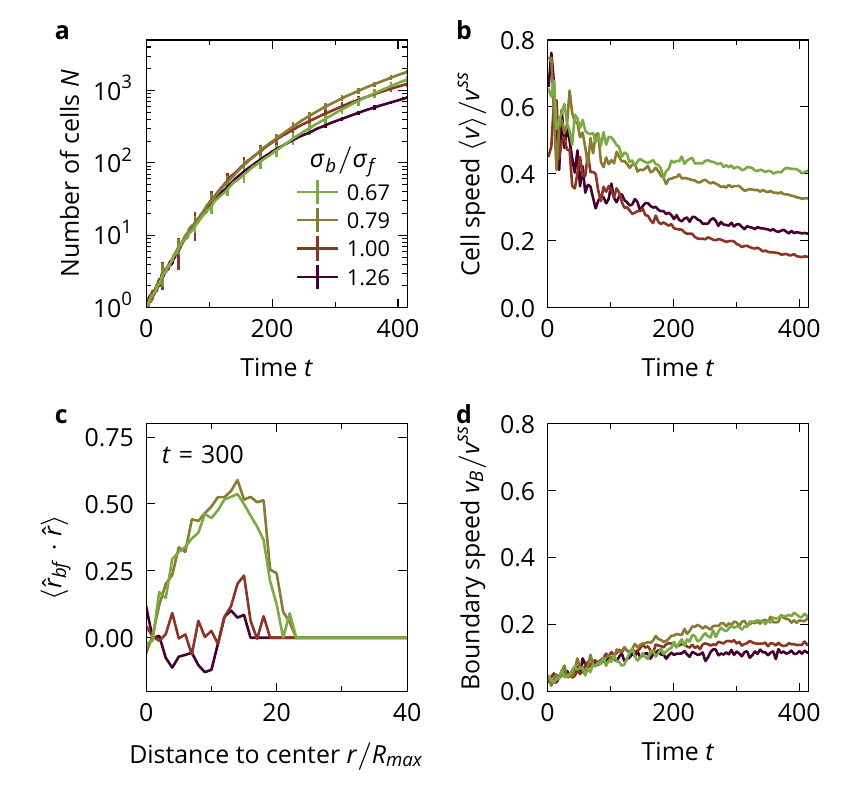}
  \caption{\textbf{Colony growth for different cell shapes.} a) Number of cells as a function of time with error bars given by the standard deviation between independent simulation runs. b) Average cell speed over time. c) Average orientation of cells in respect to the direction pointing directly away from the colony center at time $t=300$. Positive values indicate cells pointing away from the colony, negative values indicate cells pointing towards the colony. Larger values mean that cells are better aligned and/or more extended. d) Average speed of the colony boundary.}
  \label{fig:ColonyGrowth_varyShape2}
\end{figure}

\subsection{Influence of cell shape}

It is now understood that the shape of cells can be highly variable and has a strong influence on their collective behavior~\cite{Maeda2008,Keren2008,Ziebert2012,Ohta2015,Bi2015,Tjhung2017,Campo2019}. In our previous work, we showed that the present model exhibits an alignment transition in coherent migration as a function of cell shape~\cite{Schnyder2017}. Cells with a small front disk (as compared to the back disk) are bad at moving apart after colliding, while cells with a large front disk push each other out of their path which over time leads to global alignment. In this sense, cells with a large disk in our model exhibit another form of contact inhibition.

We investigate the interplay of this transition with colony growth, see \cref{fig:ColonyGrowth_varyShape2}. The exponential growth regime is found to be independent of shape, see \cref{fig:ColonyGrowth_varyShape2}a), while at long times colonies with large-front cells tend to expand more rapidly. Colonies for $\sigma_b/\sigma_f = 0.67$ are about 3 times larger than for $\sigma_b/\sigma_f = 1.26$ at the end of a simulation.
For all times, cells with larger fronts tend to be faster, see \cref{fig:ColonyGrowth_varyShape2}b). This is because cells with larger fronts are better at aligning away from the colony, as the average orientation of the cells in respect to the colony $\av{\hat r_{bf} \cdot \hat r}$ shows, \cref{fig:ColonyGrowth_varyShape2}c). Cells with small fronts tend to be slightly aligned towards the colony center. If the cells have large fronts, they tend to be aligned away from the colony, with alignment getting more pronounced towards the boundary of the colony.
In the growth of a colony, the alignment of the cells due to shape acts as a mechanism for orienting the cells away from the colony, which is what is expected from cells exhibiting contact inhibition of locomotion \cite{Stramer2016}.
As a result, the colonies of cells with large fronts expand much more quickly in the subexponential regime, see \cref{fig:ColonyGrowth_varyShape2}d) with the boundary speed being about twice as large for $\sigma_b/\sigma_f = 0.67$ than for $\sigma_b/\sigma_f = 1.26$.
The described effects are more pronounced for non-adhesive cells, see Fig.~S1.

\section{Discussion}

We investigated the dynamics of a colony of crawling, proliferating cells with a minimal, mechanical cell model. 
With a simple mechanism for contact inhibition of proliferation, we find the typical regimes of colony growth, with exponential growth at short times turning into subexponential growth at long times. The latter regime is characterised by the colony boundary moving outwards with a constant speed. We identify simple scaling relations for both regimes and the crossover between them. 

We find that the crossover colony size $A^*$ between the two regimes is a marker of contact inhibition of proliferation and that the boundary speed $v_B$ in the subexponential growth regime expresses the strength of contact inhibition of locomotion. From \cref{eq:Astar}, we see that $A^*$ depends on the steady state speed of the cells $\VSS$ and the duration of the cell cycle $T$. The faster cells the cells can migrate and the longer the cell cycle is, i.e. the longer the time between division attempts, the larger the colony can become before CIP sets in. The long-time behavior of $v_B$ only depends on the speed at which cells are able to travel, and thus primarily measures contact inhibition of locomotion, see \cref{eq:longTimeScaling}. This is corroborated by the investigation of cell shape on colony dynamics. There, the crossover $A^*$ is unchanged by cell shape, but large-front cells are better at aligning away from the colony, and as a result, tend to be more mobile and tend to expand the colony faster at long times.  It remains to be seen how close the observed faster expansion for large-front cells in the presented model is to contact enhancement of locomotion \cite{DAlessandro2017}.

The mechanisms of CIL and CIP as exhibited by our model are idealised compared to the experimental situation. While in the experiment the bulk of cells is still mobile, our cells arrest more quickly, see radial analysis. 
Similarly, our CIP response is quite strong, leading to a quicker transition to subexponential growth as in the experiment 
Also, while the mechanism of contact inhibition of locomotion -- which arises from the motile force being coupled to the cell polarisation as well as from the shape of the cell itself -- controls alignment of the cells away from the colony and colony growth, our cells are unable to expand the colony at their full crawling speed. We also neglected the polydispersity of cells sizes and noise.

Finally we would like to note that the two growth regimes are also found in other models in which the growth rate is not constant but coupled to a local property of the cells, such as density in our case. For instance, \citet{Blanch-Mercader2014} find a crossover from exponential growth to constant boundary speed in a continuum model in which the growth rate depends on the local stress; and \citet{Puliafito2012} use a one-dimensional vertex model in which cells grow if and only if stretched by neighbors and divide when reaching a critical size to qualitatively explain their data. 

\section{Methods}

\subsection{Parameters}

Introducing cell-cell adhesions made it necessary to modify the model's parameters with respect to Ref. \citenum{Schnyder2017}. We needed to shorten the maximum extension of the cells so that the disks could not be pulled apart far enough to leave a space in-between. The changed parameters also yielded cells that more readily expand from their symmetric, circular state.  We confirmed that modifying the parameters did not qualitatively change the dynamics discussed in Ref.~\citenum{Schnyder2017}.

All of the results reported here are for $\sigma_f = 1.10 R_\text{max}$ and $\sigma_b = 0.87R_\text{max}$, for a shape ratio of $\sigma_b/\sigma_f = 0.79$, except where otherwise stated. In the steady state, the cells have area $A_s = 1.44 \Rmax^2$. The division threshold is set to $R_\text{div} = 0.85 \Rmax$.

To account for the new cell-cell interactions, the energy scale is now set by $\eps_\text{core}$. We will discuss two cases, one with non-adhesive cells with $\eps_\text{well} = 0$, and one with adhesive cells, with a well depth of $\eps_\text{well} = 0.25 \eps_\text{core}$. 

Further parameter choices are $\kappa = 0.1094\, \eps_\text{core}/\Rmax^2$ and $m = 0.5\, \eps_\text{core}/R_\text{max}^2$, such that $m = 4.57 \kappa$. This yields a steady-state distance of $\rss = 0.75 \Rmax$ and steady-state speed $\VSS = \rss m / (2\zeta) = 0.19\, \eps_\text{core} / (\zeta \Rmax)$. 

All reported times are in units of $\zeta \Rmax^2/\eps_\text{core}$. The characteristic time scale of migration is $\tau_\text{mig} = \Rmax/\VSS = 5.33$. The time step of the simulations is $2.1 \cdot 10^{-3}$.

The simulations are initialized with a single cell in one of the states randomly.
We average our results over 10 individual simulation runs with different seeds for the pseudo-random number generator for the calculation of the state durations and randomisation of cell orientation after a cell division.

\subsection{Cell area}

The area $A$ of a cell with a back particle of diameter $\sigma_b$, a front particle of diameter $\sigma_f$ and a distance $r_{bf}$ between the particles is given by
\begin{align}
  \label{eq:cellArea}
	A_\text{cell} &= A_b + A_f - \text{overlap}\\
  \nonumber
	& = \pi (\sigma_b^2 + \sigma_f^2)/4\\
  \nonumber
	&+ \frac{1}{2} \sqrt{(-r_{bf} + \sigma_b + \sigma_f)(r_{bf} + \sigma_b + \sigma_f)}\\
  \nonumber
	&\times \sqrt{(r_{bf} + \sigma_b - \sigma_f)(r_{bf} - \sigma_b + \sigma_f)}
\end{align}

\begin{figure*}
  \centering
    \includegraphics[width=0.9\textwidth]{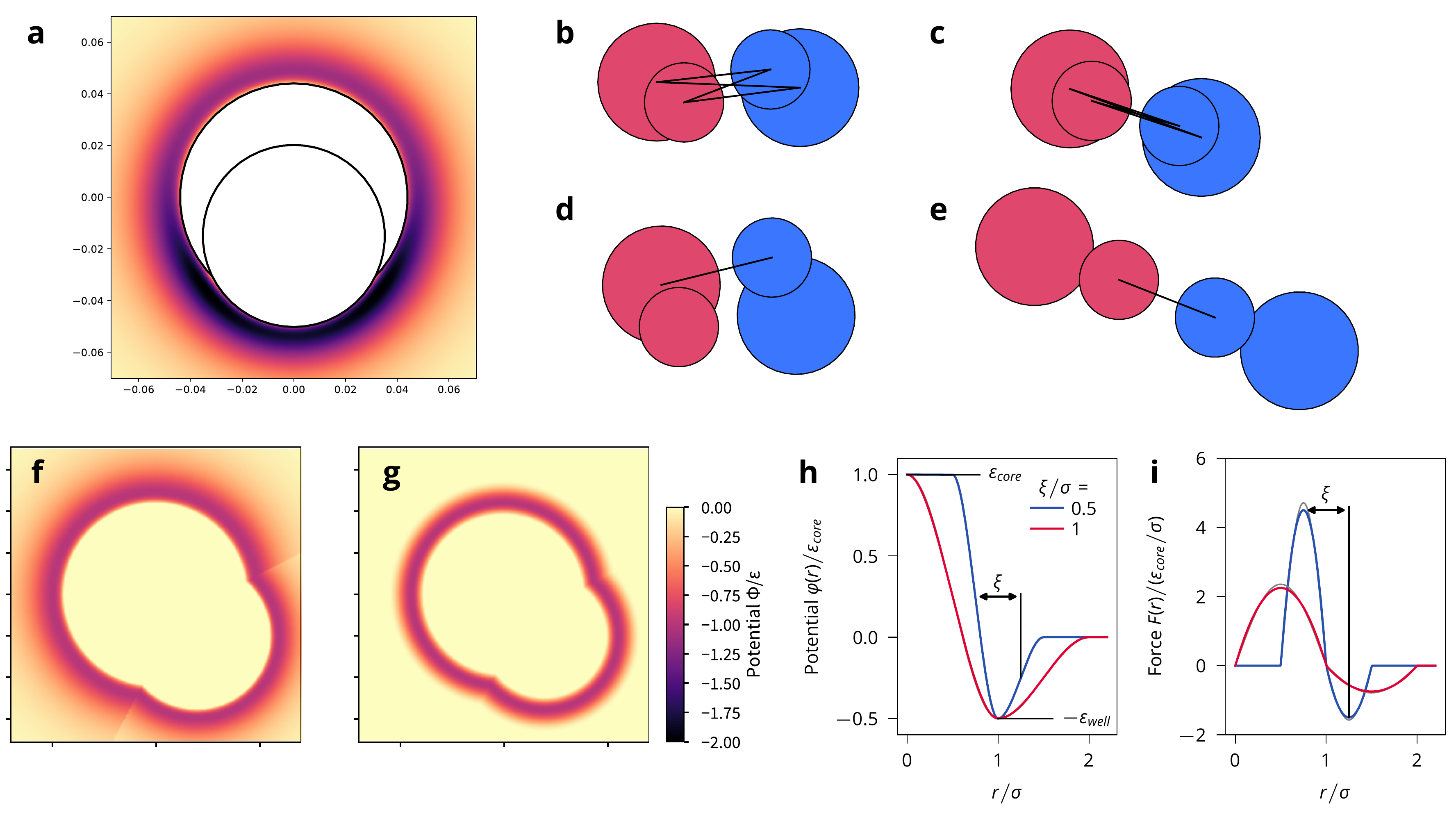}
  \caption{\textbf{Development of a soft-core potential.} 
  a) Lennard-Jones potential exerted by a cell on the disk of another cell. Note the potential minimum at the back of the smaller disk. 
  b) Illustration of the 4 possible interactions between two cells. 
  c) For two cells in contact, usually all four disks will adhere to each other, which unrealistically contracts both cells. 
  d) Illustration of allowing only the interaction between the two elements whose surfaces are closest. Analog to (b). 
  (e) Illustration of allowing only the interaction between the two elements whose surfaces are closest. Analog to (c). 
  (f) Potential landscape for Lennard-Jones potential around a cell in which only the closer surface is considered.
  (g) Potential landscape for the new softy potential around a cell in which only the closer surface is considered.
  (h) Potential and (i) force for soft-core particles with a radius of $\sigma$, energy scale $\eps_\text{core}$, with an attractive well of depth $\eps_\text{well} = \eps_\text{core}/2$, and two different well widths $\xi$. 
  }
  \label{fig:new_potential}
\end{figure*}

\subsection{Details on cell-cell interactions}

\label{section:DetailscellCellInteractions}
When two cells adhere to each other by the four possible interactions between the disks, 
 potential minima can form due to superposition of the elements' potentials at certain positions along the boundary of the cells. Those can be either behind the smaller cell element or on the sides of the cell, see \cref{fig:new_potential}a). This can lead to alignment of cells purely due to the potential, which we want to avoid. Additionally, cell elements can interact with other cells' elements with which they are not in contact and even through the other cell element, since the well depth of the potential is comparable to the range of the potential, see \cref{fig:new_potential}b). This can cause cells which are in contact to contract each other, see \cref{fig:new_potential}c). In contrast, real cell interactions occur due to direct contact of cell membranes. Therefore, to accurately model cell-cell interactions we must find a way to only have surface-surface interactions as well.

We implement surface-surface interaction by letting two cells only have up to one adhesive interaction, between the two elements whose surfaces are closest, see \cref{fig:new_potential}(d,e).
To keep the model similar to our previous paper, we still let all the cell elements repel each other. 

Finally, using the Lennard-Jones potential had one additional drawback. The width of the potential well scales with the elements' radii, which leads to a discontinuity when only the closest surface is considered adhesive and the elements have different sizes, see \cref{fig:new_potential}f). This was not an issue to our previous work in which the cutoff was set to make the forces purely repulsive, which makes the interaction very short ranged. 

We solved these issues by switching to a soft-core potential, in which the width of the well can be set independently from the element radius. We first define a cubic helper function $t(y)$ to serve as a continuous step. The function describes a continuous and monotonous step between $t(0) = 0$ and $t(\xi) = 1$, in which the end points are saddle points
\begin{align}
  t(y) &= \frac{y^2}{\xi^3}(3\xi-2y)\\
  t^{\prime}(y) &= \frac{6 y}{\xi^3}\left(\xi-y \right)
\end{align}
With defining $y = r-\sigma$,
the force and the potential between two cell disks at separation $\vec r$ are given by 
\begin{align}
  \vec F_{sc}(\vec r) = 
  \begin{cases}
    0, & y \leq - \xi \\
    (\eps_\text{core} + \eps_\text{well}) t'(-y)\hat r, &- \xi \leq y < 0\\
    - \eps_\text{well}  t'(y) \hat r, &0 \leq y < \xi \\
    0, & \xi \leq y
  \end{cases}
  \label{eq:softcore_force}
\end{align}
and 
\begin{align}
  \phi_{sc}(r) = 
  \begin{cases}
    \eps_\text{core}, & y \leq - \xi \\
    (\eps_\text{core} + \eps_\text{well})t(-y) - \eps_\text{well}, &- \xi \leq y < 0\\
     \eps_\text{well}t(y)  - \eps_\text{well}, &0 \leq y < \xi \\
     0, & \xi \leq y
  \end{cases}  
  \label{eq:softcore_potential}
\end{align}

The minimum of the potential is at $r=\sigma$, identifying $\sigma$ as the characteristic length scale of the potential, see \cref{fig:new_potential}h). $\eps_\text{well}$ is the energy in the minimum, i.e. $\phi(\sigma) = -\eps_\text{well}$. The distance between the two inflection points of the potential (the extremes of the force) is given by $\xi$, which is therefore a measure of the width of the potential well. We require $ 0 \leq \xi \leq \sigma$ to enforce that $\eps_\text{core}$ is the energy for $r = 0$, i.e. $\phi(0) = \eps_\text{core}$, and that $F(0) = 0$. The natural cutoff of the potential is $r\cut = \sigma+\xi$.

Consider now two cells with elements $\alpha$ and $\beta$, respectively ($\alpha, \beta \in [b, f]$). The separation between any pair of elements 
$(\alpha, \beta)$ is given by $\vec r_{\alpha\beta}= \vec r_\beta - \vec r_\alpha$ with unit vector $\hat r_{\alpha\beta} = \vec r_{\alpha\beta}/|\vec r_{\alpha\beta}|$.
 Their surface-to-surface distance is given by $d_{\alpha\beta} = |\vec r_{\alpha\beta}| - \sigma_{\alpha\beta}/2$ (with $\sigma_{\alpha\beta} = (\sigma_\alpha + \sigma_\beta)/2$) and the elements with the shortest surface-to-surface distance are denoted $(\alpha', \beta')$, i.e.
 \begin{align}
   d_{\alpha'\beta'} = \min_{\alpha,\beta}(d_{\alpha\beta}).
 \end{align}
Then, the (cell-cell) force $\vec F_{cc,\alpha}$ acting on cell element $\alpha$ exerted by the elements of the other cell is given by
\begin{align}
  \vec F_{cc,\alpha} &= \sum_{\beta} \vec F_{\alpha\beta}\\
  \vec F_{\alpha\beta} &= 
  \begin{cases}
    \vec F_{sc}(\vec r_{\alpha\beta}) &\text{if }  \vec F_{sc}(\vec r_{\alpha\beta})\hat r_{\alpha\beta} < 0 \text{ or } (\alpha, \beta) = (\alpha', \beta')\\
    0 &\text{otherwise}
  \end{cases}
\end{align}
using \cref{eq:softcore_force} for $\vec F_{sc}(\vec r_{\alpha\beta})$.

With this soft-core potential, the potential well around a cell is now of constant width and of shorter range, see \cref{fig:new_potential}g). For all our simulations we chose a width of the potential well of $\xi = 0.654 \Rmax$ ($= 0.75\sigma_b$).

\subsection{Radial distribution of successful cell divisions}

We calculated the radial distribution of successful cell divisions for all times, by assuming that the colony is circular, i.e. radially symmetric. We obtain 
\begin{align}
 &N_\text{div}(r, t) = \\
 \nonumber
 &\frac{1}{A_\text{ring}(r)} \frac{1}{\Delta t} \int_{t-\Delta t}^t \int_{r-\Delta r, r} \sum_i \delta(\vec r\,' - \vec R_{i,\text{new}}(t)) dr'dt
\end{align}
with the center of mass of the colony placed at the origin, $A_\text{ring}(r) = \pi[r^2 - (r-\Delta r)^2]$, and $R_{i,\text{new}}(t))$ the position of the $i$-th division event occurring at time $t$. 

\subsection{Average cell orientation}

We quantified the orientation of the cells in respect to the colony as follows. With $\vec r$ the position of the cell in respect to the center of mass of the colony, and $\vec r_{bf}$ the cell's extension, and $\hat r$ and $\hat r_{bf}$ denoting their unit vectors, we measure the orientation of cells towards or away from the colony
\begin{align}
  p(r) = \av{\hat r_{bf} \cdot \hat r}
\end{align}
 as a function of $r = |\vec r|$. This function yields values between $-1$ and $1$. If cells on a ring of radius $r$ are mostly pointing towards the inside of the colony, $p(r)< 0$, whereas if the cells are mostly pointing away from the colony, then $p(r)>0$. 
 
\section{Acknowledgements}

We thank Carles Blanch-Mercader, Joseph d'Alessandro, Jens Elgeti, Norihiro Oyama, Mitsusuke Tarama, and Pascal Silberzan for helpful and insightful discussions. 
\textbf{Funding:} We acknowledge support by the Japan Society for the Promotion of Science (JSPS) KAKENHI Grants No. 26610131, 16H00765, and 17H01083. \textbf{Competing interests:} The authors declare that they have no competing interests.
\textbf{Author contributions:}
SKS and RY designed research. SKS and JJM worked on the simulations. SKS performed the simulations and analyzed the data. All authors contributed to interpreting the results and writing the paper.
\textbf{ Data and materials availability: }All data needed to evaluate the conclusions in the paper are present in the paper and/or the Supplementary Materials. Additional data related to this paper may be requested from the authors.

\end{document}